\documentclass[journal=jacsat,manuscript=article]{achemso}

\usepackage{chemformula}
\usepackage[T1]{fontenc}
\usepackage{url}

\author{Daniel Graf}
\email{dg641@cam.ac.uk}
\author{Alex J. W. Thom}
\affiliation[University of Cambridge]
{Yusuf Hamied Department of Chemistry, University of Cambridge, Cambridge}

\title[C(HF)-RPA]
  {Corrected Density Functional Theory and the Random Phase Approximation:
  Improved Accuracy at Little Extra Cost}


\begin{document}

\begin{tocentry}
    toc
\end{tocentry}

\begin{abstract}
    We recently introduced an efficient methodology to perform
    density-corrected Hartree--Fock density functional theory (DC(HF)-DFT)
    calculations and an extension to it we called ``corrected'' HF DFT
    (C(HF)-DFT). 
    In this work, we take a further step and combine C(HF)-DFT, 
    augmented with a straightforward orbital energy correction, with the 
    random phase approximation (RPA). We refer to the resulting methodology
    as corrected HF RPA (C(HF)-RPA).
    We evaluate the proposed methodology across various RPA methods: direct RPA
    (dRPA), RPA with an approximate exchange kernel (RPA-AXK), and RPA with
    second-order screened exchange (RPA-SOSEX).
    C(HF)-dRPA, in particular, demonstrates very promising performance; 
    for RPA with exchange methods we find over-corrections for certain
    chemical problems.
\end{abstract}

Density functional theory (DFT) can undoubtedly be considered a highly
successful theory and a major driving force in computational chemistry,
physics, and materials science. However, despite its success, it is
well-known that standard density functional approximations (DFAs) are incapable
of accurately describing dispersion interactions.\cite{becke1995}
Various
approaches, such as Grimme's dispersion corrections,\cite{grimme2016a,
grimme2004a, grimme2006a, grimme2006b, grimme2010a, grimme2017a, grimme2019a} 
have been developed to
address this limitation. While incorporating corrections obtained
from stand-alone methods has proven to be a valid approach with widespread use
and success, we believe that an electronic structure method containing an
intrinsic description of dispersion is even more appealing.

One such method that possesses this desirable property is the random phase
approximation (RPA),\cite{bohm1951a, pines1952a, bohm1953a}
which, as an adiabatic-connection method,\cite{langreth1975a, langreth1977a}
can be
seen as sitting on the border between DFT and wave-function theory.
In addition to its ability to accurately describe dispersion, RPA is
size-consistent,\cite{fuchs2005a} 
applicable to small gap systems (contrary to e.g. 
M\o{}ller--Plesset Perturbation theory of second order),\cite{fuchs2005a,
harl2008a, harl2009a} 
and can be implemented in a highly efficient, linear-scaling
fashion.\cite{schurkus2016a, luenser2017a, graf2018a, kallay2015a}  
Furthermore, there exists a clearly defined, albeit extremely expensive, 
route towards exactness, setting it apart from standard
DFAs.\cite{goerling2019a}

While self-consistent versions of RPA have been presented in the
literature,\cite{hellgren2012a, voora2019a, yu2021a, graf2019a, graf2020a}
RPA is commonly employed in a post-Kohn--Sham fashion,\cite{furche2001a,
eshuis2010a} 
utilising orbitals and
orbital energies from a preceding DFA calculation, which we will refer to as
the ``reference calculation''. Most commonly, the reference
calculation is performed using a generalized gradient approximation (GGA), with
the one proposed by Perdew, Burke, and Ernzerhof (PBE)\cite{perdew1996,
perdew1997} being particularly 
popular. 
Considering the ever-increasing demand for highly efficient yet
accurate methods, it makes sense to combine the computational efficiency of
modern RPA implementations with a cheap self-consistent field calculation.
However, it is important to note that pure density functionals are known for
their self-interaction error and the resulting over-delocalisation of charge.
\cite{sanchez2008, cohen2007, cohen2008, li2017, li2015, johnson2013, 
vazquez2015, leblanc2018, perdew1981, sanchez2006, vydrov2007, ruzsinszky2007,
sanchez2009, yang2012, perdew1982, zhang1998, ruzsinszky2006, cohen2014}
Consequently, these issues can lead to erroneous densities,
Kohn--Sham (KS) orbitals, and orbital energies, which are subsequently used as
input for the RPA calculation.

Previous research has demonstrated that evaluating the density functional on
the Hartree--Fock (HF) density instead of the self-consistent one significantly
improves accuracy in many cases.\cite{gordon1972, colle1975, scuseria1992, 
janesko2008, cioslowski1993, oliphant1994, verma2012, santra2021}
These findings have led Burke and
co-workers to develop the density-corrected Hartree--Fock density functional
theory (DC(HF)-DFT) framework, where the self-consistent DFA density is
replaced by the HF density \textit{if} the DFA density is found to be 
erroneous.\cite{nam2021, nam2020, sim2018,
sim2022, vuckovic2019, song2021, kim2013, martin2021, kim2019, kim2014,
wasserman2017}

We recently proposed a simple heuristic to determine whether the
self-consistent DFA density should be replaced by the HF
density.\cite{graf2023a}
The key idea is to examine the behaviour of the non-interacting kinetic energy,
which should decrease in magnitude if the density functional over-delocalises
charge. To detect this, we compare the non-interacting kinetic energy obtained
from the converged DFA calculation with the one obtained from a converged HF
calculation. If the HF non-interacting kinetic energy is larger than the DFA
one, we can conclude that the HF density is a better choice. This can be
quantified by the relative change in the non-interacting kinetic energy, given
by
\begin{equation}
    r_{\text{kin}} = \frac{T^{\text{HF}}_{\text{s}} - T^{\text{KS}}_{\text{s}}}
    {T^{\text{KS}}_{\text{s}}},
\end{equation}
where
\begin{equation}
    T_{\text{s}} = - \frac{1}{2} \sum^{N_{\text{occ}}}_{i} 
    \int\!\mathrm{d} \mathbf{r} \:\phi^{*}_{i} ( \mathbf{r} ) \nabla^{2}_{1} 
    \phi_{i} ( \mathbf{r} ).
\end{equation}
So, if $r_{\text{kin}}$ is positive then the HF density should be used.

Converging a HF calculation can be computationally expensive. Therefore, we
proposed a more efficient procedure, which involves the following steps:
\begin{enumerate}
    \item Converge the DFA calculation.
    \item Evaluate the Fock matrix $\mathbf{F}$ using the converged DFA
        one-particle density matrix $\mathbf{P}$.
    \item Update the orbitals once.
    \item Evaluate $T_{\text{s}}$ using the updated orbitals.
    \item Calculate $r_{\text{kin}}$.
\end{enumerate}
As mentioned earlier, the RPA can be derived within the adiabatic-connection
formalism, where all parts of the energy except for the correlation energy are
treated exactly.\cite{langreth1975a, langreth1977a}
The total RPA energy is given by
\begin{equation}
    E^{\text{RPA}} = 
    \underbrace{E_{\text{h}}\left[\phi^{\text{KS}}\right] 
    + E_{\text{J}}\left[\phi^{\text{KS}}\right] 
    + E_{\text{X}}\left[\phi^{\text{KS}}\right]}_{E^{\text{HF}}  
    \left[\phi^{\text{KS}}\right]}
    + E^{\text{RPA}}_{\text{c}}
    \left[\phi^{\text{KS}}, \epsilon^{\text{KS}}\right] ,
\end{equation}
where $E_{\text{h}}$, $E_{\text{J}}$, and $E_{\text{X}}$ denote the 
one-electron, the classical Coulomb, and the exact exchange energy, 
respectively. 
It is important to note that the first three terms are equivalent to evaluating
the HF expression using KS orbitals. Due to the resulting requirement of 
constructing a Fock matrix for evaluating the total RPA energy, 
the RPA method aligns remarkably well with our proposed DC(HF)-DFT procedure.

The RPA correlation energy depends not only on the KS orbitals,
which determine the DFA density, but also on the corresponding orbital
energies. Errors in the KS potentials can --- and will --- affect the orbital
energies,\cite{kuemmel2008} 
introducing additional sources of error in the total RPA energy. 
Yang and colleagues have recently proposed a rigorous method to correct 
orbital energies,\cite{zheng2011a, mei2021a, li2015, li2017, su2020a,
mei2020a, mei2022a}
albeit at a significant computational cost.
Alternatively, Ochsenfeld and colleagues\cite{lemke2022a}
have presented a more computationally 
efficient scheme to correct orbital energies obtained from a GGA calculation 
by diagonalising a projected KS matrix\cite{voora2019a, graf2019a, graf2020a}
\begin{equation}
    \mathbf{\tilde{H}} [\mathbf{P}^{\text{GGA}}] = 
    \mathbf{S} \mathbf{P}^{\text{GGA}}
    \mathbf{H}^{\text{HGGA}} [\mathbf{P}^{\text{GGA}}]
    \mathbf{P}^{\text{GGA}} \mathbf{S} 
  + \mathbf{S} \mathbf{P}^{\text{virt, GGA}}
    \mathbf{H}^{\text{HGGA}} [\mathbf{P}^{\text{GGA}}]
    \mathbf{P}^{\text{virt, GGA}} \mathbf{S} ,
    \label{eq:semproj}
\end{equation}
where $\mathbf{P}^{\text{virt, GGA}}$ represents the virtual one-particle
density matrix.    
The projection ensures that the
post-diagonalisation orbitals reproduce the same one-particle density matrix as
the one used to construct the KS matrix, allowing for the reuse of the 
evaluated exact exchange matrix --- contained in the hybrid one-particle 
Hamiltonian $\mathbf{H}^{\text{HGGA}}$ --- in the calculation of the 
RPA energy.

\begin{figure}
\begin{tikzpicture}[
  node distance=1.25cm,
  box/.style={
    draw,
    text width=3cm,
    align=center,
    font=\footnotesize
  },
  oval/.style={
    draw,
    rounded corners=0.2cm,
    text width=2cm,
    align=center,
    font=\footnotesize
  },
  arrow/.style={
    ->,
    >=latex
  }
]

\node[box] (step1) {Converge GGA};
\node[box, below of=step1] (step2) {Evaluate 
    $\mathbf{F}[\mathbf{P}^{\text{GGA}}]$};
\node[box, below of=step2] (step3) {Update orbitals};
\node[box, below of=step3] (step4) {Evaluate $r_{\text{kin}}$};
\node[oval, below of=step4] (step5) {$r_{\text{kin}} < 0$?};
\node[box, below left of=step5, xshift=-1.25cm] (step6) {Evaluate 
    $\mathbf{\tilde{H}}[\mathbf{P}^{\text{GGA}}]$};
\node[box, below of=step6] (step7) {Correct $\epsilon^{\text{KS}}$};
\node[box, below right of=step5, xshift=1.25cm] (step8) {Converge HF};
\node[box, below of=step5, yshift=-2cm] (step9) {Evaluate $E^{\text{RPA}}$};

\draw[arrow] (step1) -- (step2);
\draw[arrow] (step2) -- (step3);
\draw[arrow] (step3) -- (step4);
\draw[arrow] (step4) -- (step5);
\draw[arrow] (step5.west) -| node[midway, left] {Yes} (step6.north);
\draw[arrow] (step6) -- (step7);
\draw[arrow] (step5.east) -| node[midway, right] {No} (step8.north);
\draw[arrow] (step7.south) |- (step9.west);
\draw[arrow] (step8.south) |- (step9.east);
\draw[arrow] (step2.west) -- node[above] {Reuse} ++(-2.5cm,0) |-  (step6.west);
\draw[arrow] (step2.west) -- ++(-2.5cm,0) |- (step9.west);
\end{tikzpicture}
\caption{Schematic representation of the C(HF)-RPA approach.}
\label{fig:crpa_procedure}
\end{figure}
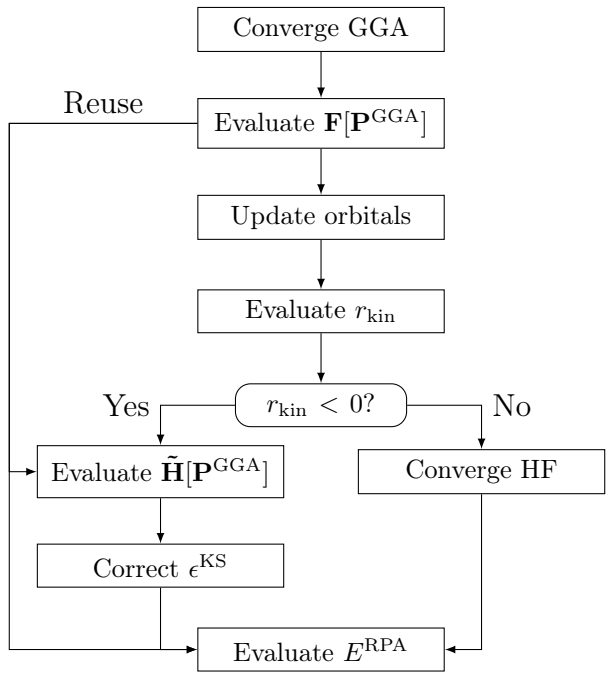

By integrating this method of orbital correction with our previously proposed
DC(HF)-DFT procedure, we introduce the corrected HF RPA (C(HF)-RPA) approach, 
as depicted in Figure~\ref{fig:crpa_procedure}.

To evaluate the performance of our proposed procedure, we conducted tests on
diverse sets of chemical problems using the direct RPA (dRPA), RPA with 
an approximate exchange kernel (RPA-AXK),\cite{bates2013a, beuerle2018a} 
and RPA with second-order screened exchange (RPA-SOSEX)\cite{grueneis2009a,
freeman1977a, beuerle2018a} methods. 
Table \ref{tbl:summary_test_sets} provides an overview of the test
sets investigated in this study.
\begin{table}
  \caption{Summary of the test sets investigated in this work.}
  \label{tbl:summary_test_sets}
  {\small
  \begin{tabular}{ll}
    \hline
      Test set & Description \\
    \hline                                              
      S22\cite{jurecka2006, marshall2011} & Non-covalent interaction energies 
                 (hydrogen bonds, weak dispersion bonds, mixed) \\
      B30\cite{bauza2013, roza2014} & Non-covalent interaction energies 
                 (halogen, chalcogen, and pnicogen bonds) \\
      FH51\cite{friedrich2013} & Reaction energies in various (in-)organic 
                 systems \\
      DARC\cite{goerigk2010, goerigk2017, johnson2008} & Reaction energies of 
                 Diels-Alder reactions \\
      G21EA\cite{curtiss1991, goerigk2010, goerigk2017} & Adiabatic electron 
                 affinities \\
      G21IP\cite{curtiss1991} & Adiabatic ionisation potentials \\
      SIE4x4\cite{goerigk2010} & Self-interaction-error related problems \\
      W4-17\cite{karton2017a} & Total atomisation energies \\
    \hline
  \end{tabular}
  }
\end{table}
The results obtained using the different approaches are presented in 
Tables~\ref{tbl:maes_normal_cases} and~\ref{tbl:maes_challenging_cases}. 
Additionally, we include the results obtained using the widely used hybrid 
functional B3LYP\cite{becke1988a, lee1988a, becke1993a}
for comparison, as its computational cost is comparable to 
evaluating dRPA on top of a GGA calculation using, for instance, the PBE 
functional.

Starting with Table~\ref{tbl:maes_normal_cases}, it is evident that the 
C(HF)-RPA approach significantly improves upon the results obtained with 
standard RPA approaches.
Notably, the improvement for the B30 test set primarily arises from
correcting the GGA density, while the improvement for the DARC test set is
attributed to the orbital energy correction. Previous studies have already
highlighted the challenges posed by the B30 test set for standard local and
semi-local DFAs, demonstrating that these functionals exhibit significant
density-driven errors for this test set.\cite{sim2022}

Furthermore, it is worth mentioning the considerably better performance of all 
RPA methods compared to B3LYP. 
The substantial discrepancy observed here stems from the
inherent limitations of standard DFAs in capturing dispersion interactions, as
mentioned in the introductory part of this work. While applying a dispersion
correction could partially mitigate this issue, our intention here is to
emphasise the intrinsic capabilities of RPA and its independence from separate
correction schemes.
\begin{table}
  \caption{Mean absolute errors in kcal/mol of different RPA methods and B3LYP
    for the S22, B30, FH51, and DARC test sets.}
  \label{tbl:maes_normal_cases}
  \begin{tabular}{ccccc}
    \hline
                           & S22   & B30   & FH51  & DARC    \\
    \hline                                                 
      dRPA@PBE             & 0.236 & 1.537 & 2.444 & 1.080   \\
      C(HF)-dRPA@PBE       & 0.240 & 0.653 & 1.566 & 0.301   \\
    \hline                                                 
      RPA-AXK@PBE          & 0.227 & 1.680 & 1.293 & 1.193   \\
      C(HF)-RPA-AXK@PBE    & 0.266 & 0.786 & 1.091 & 0.443   \\
    \hline                                              
      RPA-SOSEX@PBE        & 0.427 & 1.202 & 1.850 & 2.523   \\
      C(HF)-RPA-SOSEX@PBE  & 0.383 & 0.738 & 2.243 & 2.301   \\
    \hline                                              
      B3LYP                & 3.137 & 1.123 & 4.111 & 14.996  \\
    \hline
  \end{tabular}
\end{table}

Table~\ref{tbl:maes_challenging_cases} presents test sets that are particularly
challenging for dRPA due to their sensitivity to self-interaction errors. The
G21EA, G21IP, and SIE4x4 test sets have also posed difficulties for the
recently proposed $\sigma$-functionals by the G\"{o}rling group, which are
considered highly promising.\cite{trushin2021a, fauser2021a, erhard2022a}
Remarkably, the C(HF)-dRPA approach achieves
tremendous improvements for these three test sets. 
However, for C(HF)-RPA-AXK and C(HF)-RPA-SOSEX, while there is also significant 
improvement in the accuracy for the SIE4x4 test set, there is a significant 
decrease in accuracy for the G21EA and G21IP test sets.

\begin{figure}
    \includegraphics{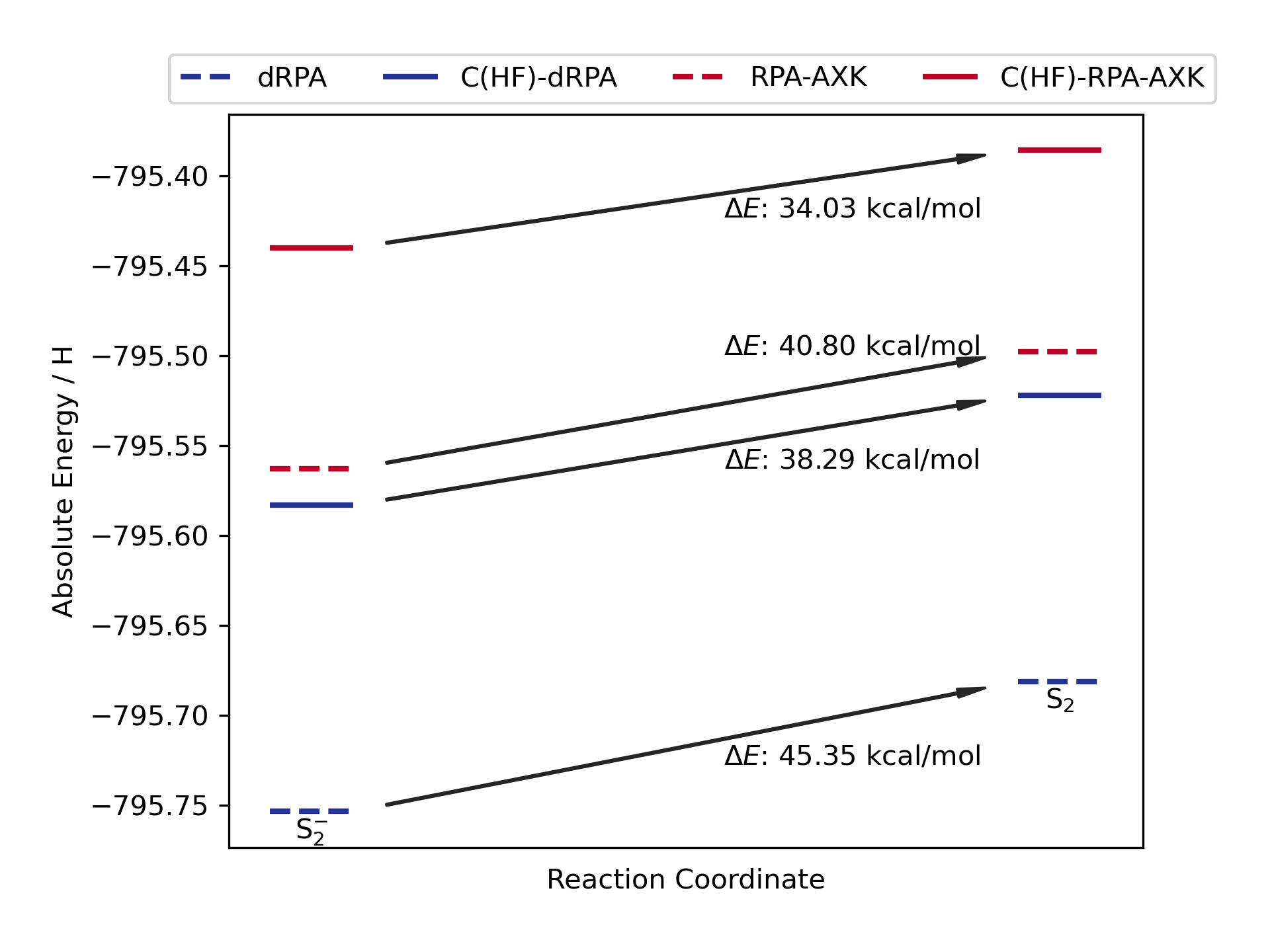}
    \caption{\label{fig:reaction_behaviour_g21ea} Example for the change in the
    absolute energies and the reaction energies for dRPA, RPA-AXK, and their
    density-corrected variants. The reference value for the reaction energy is
    38.0 kcal/mol.}
    \label{fig:example_g21ea}
\end{figure}

To shed light on this observation, we illustrate the behaviour of dRPA, 
RPA-AXK, and their corrected counterparts using an example reaction from the 
G21EA test set, as shown in Figure~\ref{fig:example_g21ea}.
For dRPA, it can be observed that the anionic reactant ($\text{S}_{2}^{-}$) is
too stable compared to the neutral product ($\text{S}_{2}$). This discrepancy
can be explained by the too deep correlation hole and the resulting
over-correlation due to the absence of Pauli repulsion between the
particle-hole pairs within dRPA, which is particularly pronounced in the system
with an additional electron. In the case of this reaction, the kinetic energy
indicator suggests using the Hartree--Fock reference instead of the
GGA reference. Therefore, not only are the
GGA orbitals replaced by the HF orbitals, but also the GGA orbital energies are
substituted with HF orbital energies.
While it is known that GGAs produce too small gaps between the highest
occupied molecular orbitals (HOMOs) and the lowest unoccupied molecular 
orbitals
(LUMOs), HF tends to produce too large HOMO-LUMO gaps. 
Since the density response,
and consequently the RPA correlation energy, is directly influenced by the
HOMO-LUMO gap, an increase in this gap leads to a decrease in the dRPA
correlation energy. The excessively large HOMO-LUMO gap resulting from HF
calculations seems to counteract the self-interaction within dRPA, leading to a
highly accurate reaction energy.
When considering the RPA-AXK approach, it can be observed that the correlation
energy is ``corrected'' twice: 
first, by the inclusion of Pauli repulsion in the
response kernel, and second, by the large HF HOMO-LUMO gap. However,
this double correction leads to an overall decrease in accuracy. The same trend
is observed for RPA-SOSEX.
\begin{table}
  \caption{Mean absolute errors in kcal/mol of different RPA methods and B3LYP 
    for the G21EA, G21IP, SIE4x4, and W4-17 test sets.}
  \label{tbl:maes_challenging_cases}
  \begin{tabular}{ccccc}
    \hline
                           & G21EA & G21IP & SIE4x4 & W4-17  \\
    \hline                                         
      dRPA@PBE             & 6.009 & 5.323 & 21.314 & 26.120 \\
      C(HF)-dRPA@PBE       & 3.158 & 3.551 & 9.608  & 36.616 \\
    \hline                                      
      RPA-AXK@PBE          & 1.992 & 1.973 & 13.530 & 16.317 \\
      C(HF)-RPA-AXK@PBE    & 5.282 & 4.325 & 3.507  & 35.705 \\
    \hline                                      
      RPA-SOSEX@PBE        & 3.477 & 2.932 & 11.704 & 13.390 \\
      C(HF)-RPA-SOSEX@PBE  & 7.339 & 5.427 & 9.218  & 39.799 \\
    \hline                                      
      B3LYP                & 2.950 & 3.751 & 17.951 & 4.687  \\
    \hline
  \end{tabular}
\end{table}

\begin{figure}
    \includegraphics{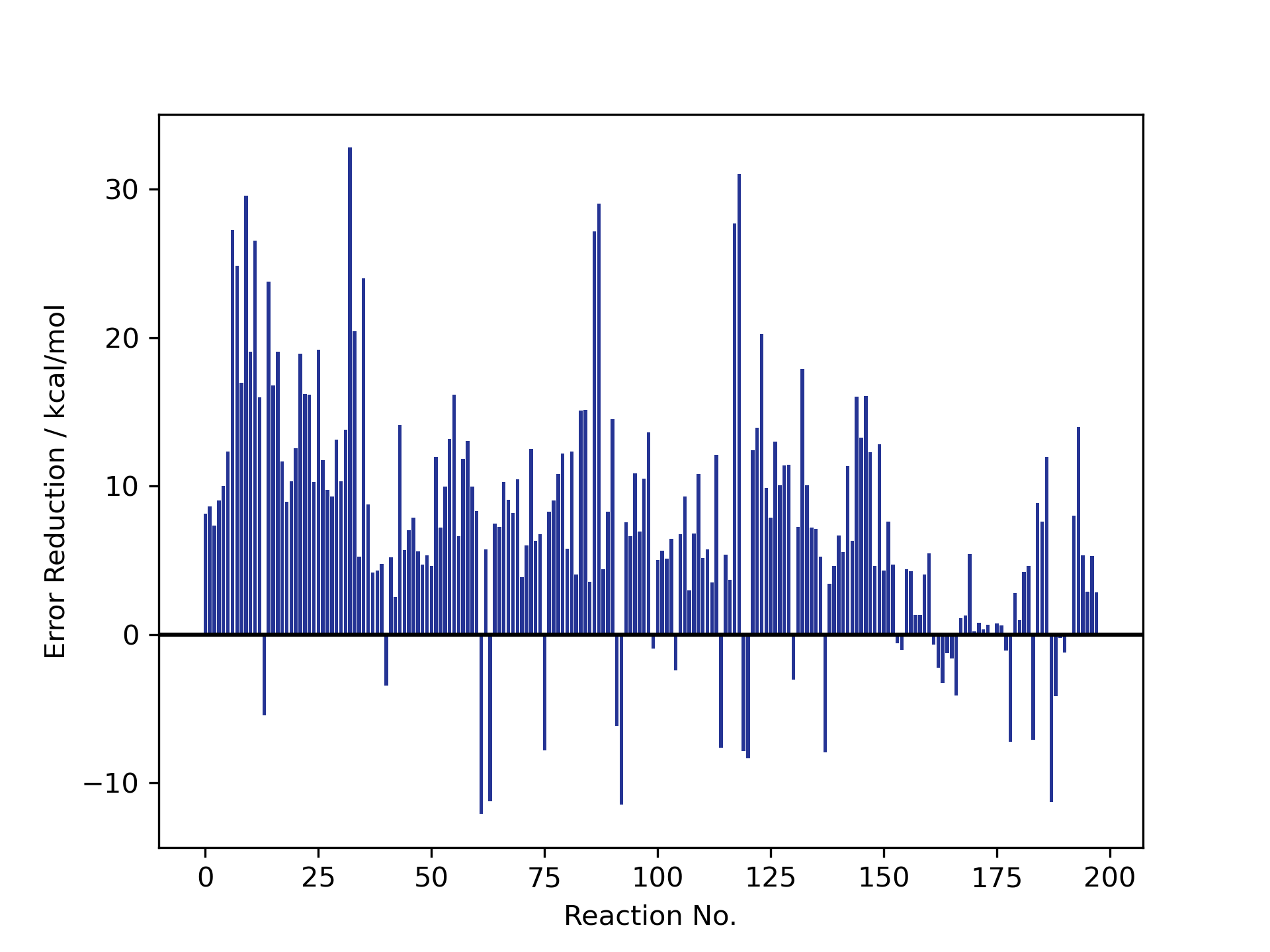}
    \caption{Error reduction when
    applying DC(HF)-DFT in combination with the kinetic energy indicator for
    the W4-17 test set for the PBE functional. Positive values indicate
    improved accuracy, negative values decreased accuracy.}
    \label{fig:dcpbe_w417}
\end{figure}

When examining the W4-17 test set, which consists of atomisation energies, it
is unfortunate to observe that the results are significantly degraded when
employing the C(HF)-RPA scheme. To investigate whether the issues arise
from the kinetic energy indicator erroneously suggesting incorrect references,
we compared the performance of standard PBE and DC(HF)-PBE using our indicator.
The reduction in errors for the various reactions in the W4-17 test set is
depicted in Figure~\ref{fig:dcpbe_w417}. 
As evident from the results, although there are some
increases in errors (values below $0$ in Figure~\ref{fig:dcpbe_w417}), the
overall performance of the kinetic energy indicator aligns with expectations:
it selects the density that leads to improved accuracy.

The reason behind the decreased performance of the RPA approaches after
correction lies in the fact that the stabilities of the bound systems are
excessively reduced compared to the individual atoms, primarily due to the
overly large HF HOMO-LUMO gaps. As RPA already tends to underestimate the
stability of bound systems, this further amplifies the errors, resulting in
larger inaccuracies.

\begin{figure}
    \includegraphics{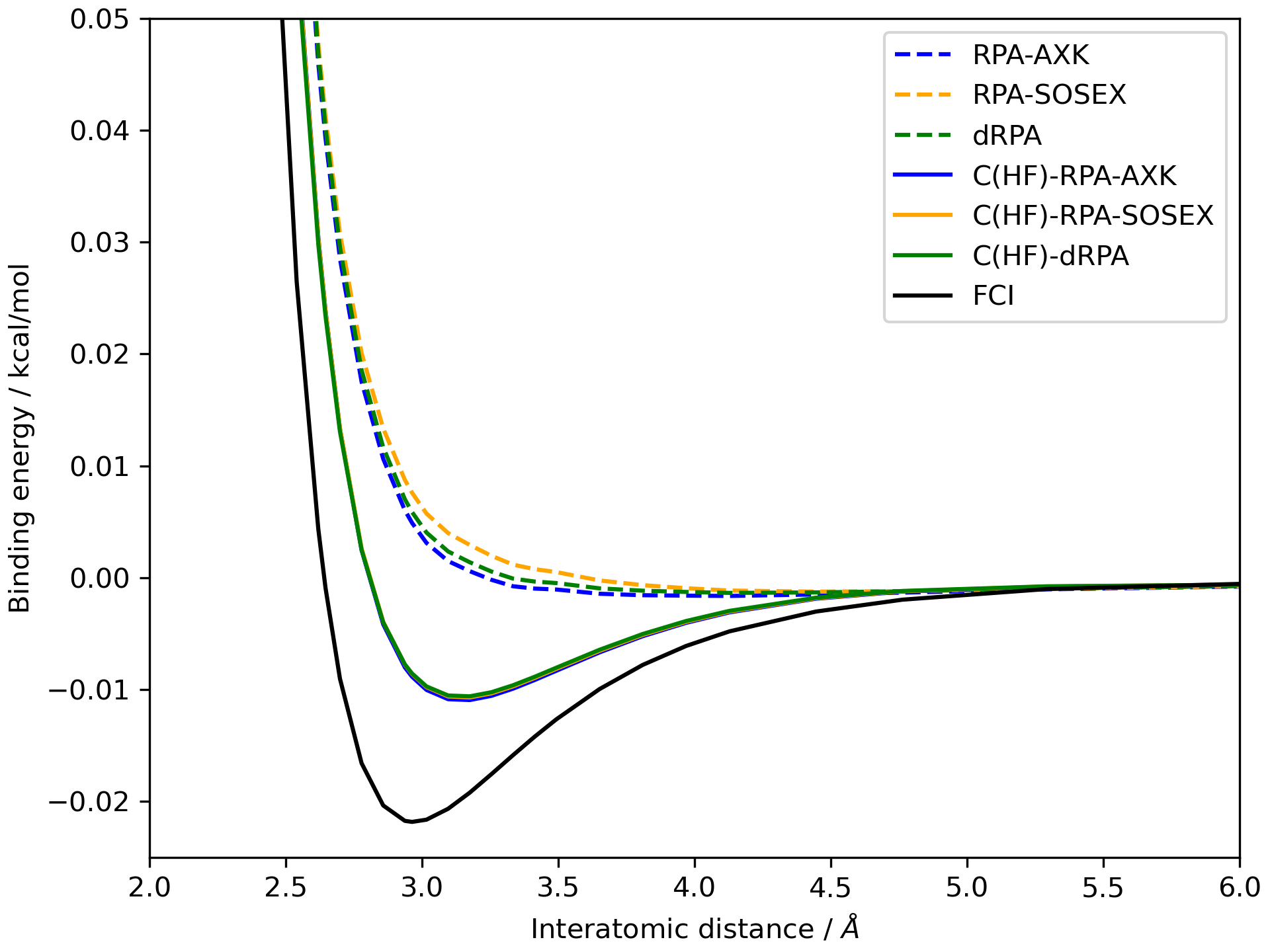}
    \caption{Dissociation of a helium dimer calculated
    with the different (corrected) RPA methods. As reference serves an
    estimated complete basis set FCI curve.\cite{mourik1999a}}
    \label{fig:he2_diss} 
\end{figure}

Finally, we calculated the dissociation curve of a helium dimer. The
results are presented in Figure~\ref{fig:he2_diss}. 

It is evident that none of
the RPA methods can generate a binding potential energy curve when performed on
top of a PBE calculation. However, when employing our proposed corrected RPA
procedure, all RPA methods yield binding curves of comparable quality.

In conclusion, our work involved the integration of the density-corrected DFT
framework, augmented by a straightforward orbital energy
correction,\cite{lemke2022a} with the
random phase approximation (RPA). This combination resulted in a novel
methodology we call corrected HF random phase approximation (C(HF)-RPA). 
Notably, C(HF)-RPA exhibits particular appeal when utilised in conjunction 
with our recently introduced kinetic energy indicator,\cite{graf2023a} as it 
enables efficient recycling of quantities necessary for the computation of the 
total RPA energies.

We demonstrated that our C(HF)-RPA approach effectively enhances the 
performance of standard RPA methods. 
Particularly noteworthy are the outcomes obtained with
C(HF)-dRPA, as it not only enhances results for non-covalent interactions and
reaction energies but also shows significant improvements in challenging
scenarios such as adiabatic electron affinities, adiabatic ionisation
potentials, and self-interaction related problems.

Combining the C(HF)-RPA scheme with RPA methods incorporating exchange can 
result in over-corrections for certain chemical problems. Therefore, it may be
advisable to limit the application of C(HF)-RPA to dRPA. 
However, considering the
findings presented in this work, this limitation should not be regarded as a
drawback. The performance of C(HF)-dRPA is comparable to that of standard RPA
methods with exchange, while offering the advantage of significantly lower
computational cost.

However, it is important to exercise caution when considering atomisation
energies, as the performance of C(HF)-RPA was notably inferior in this aspect
compared to standard RPA.
It is worth mentioning that an intriguing avenue for further exploration would
involve optimising a $\sigma$-functional\cite{trushin2021a, fauser2021a, 
erhard2022a} in conjunction with the presented
C(HF)-dRPA approach. This has the potential to yield significantly
improved results.

Finally, we would like to reiterate the remarkable potential of RPA,
particularly the recently introduced $\sigma$-functionals, as highly promising
electronic-structure methods. They offer very good performance with relatively
low computational cost, comparable to that of hybrid DFAs. Our hope is that
this work not only brings about changes and improvements in the utilisation of
RPA but also serves as a foundation for developing new and more accurate
$\sigma$-functionals.

\section{Computational details}
The calculations were performed utilising a developmental version
of the FermiONs++ software package, developed by the Ochsenfeld group
\cite{kussmann2013, kussmann2015, kussmann2017}. The software binary was
compiled using the GNU Compiler Collection (GCC) version 12.1. The computations
were carried out on a compute node equipped with 2 Intel Xeon E5-2630
v4 CPUs, featuring a total of 20 cores and 40 threads with a clock speed of 
2.20\,GHz.

The calculations of the exchange-correlation terms were conducted using the
multi-grids specified in Ref.~\citenum{laqua2018}, employing a smaller grid
during the SCF optimisation and a larger grid for the final energy evaluation.
These grids were generated using the modified Becke weighting
scheme.\cite{laqua2018} The convergence criterion for the SCF calculations was
set to $10^{-6}$ for the norm of the difference density matrix $|| \Delta
\mathbf{P} ||$.

Unless stated otherwise, we employ the integral-direct 
resolution-of-the-identity Coulomb (RI-J)  method of 
Kussmann \textit{et al.}\cite{kussmann2021} for the
evaluation of the Coulomb matrices and the linear-scaling semi-numerical exact
exchange (sn-LinK) method of Laqua \textit{et al.}\cite{laqua2020} 
for the evaluation of the exact exchange matrices.

By default, we employ the frozen-core approximation for the calculation of 
RPA correlation energies. The integration along the imaginary frequency axis is
carried out using an optimised minimax grid\cite{kaltak2014a, graf2018a}
consisting of 15 quadrature points.

For the S22 test set, we utilised the cc-pVTZ\cite{dunning1989, koput2002a,
prascher2011, wilson1999a, woon1993, woon1994, balabanov2005a} 
atomic orbital basis in
combination with the cc-pVTZ-RI\cite{weigend2002a, hill2008a, haettig2005a}
and cc-pVTZ-JKFIT\cite{weigend2002} 
auxiliary basis sets for RPA and RI-J, respectively.
For the G21EA test set, we employed the aug-cc-pVQZ\cite{dunning1989, 
kendall1992, woon1994, prascher2011, woon1993} atomic orbital 
basis along
with the corresponding auxiliary basis\cite{weigend2002a, haettig2005a,
bross2013a} for RPA, and the cc-pVTZ-JKFIT
auxiliary basis for RI-J.
In the case of the W4-17 test set, we utilised the large
aug-cc-pwCVQZ\cite{balabanov2005a, dunning1989, kendall1992, peterson2002a,
wilson1999a, woon1993} atomic
orbital basis in conjunction with the respective auxiliary basis
set\cite{haettig2005a, haettig2012a, weigend2002a} for
both RPA and RI-J; no frozen-core approximation was employed, and
reactions 9, 134, and 183 were excluded due to technical difficulties.
For the remaining test sets, we employed the cc-pVQZ\cite{balabanov2005a,
dunning1989, koput2002a, prascher2011, wilson1999a, woon1993, woon1994}
atomic orbital basis in
combination with the respective auxiliary basis for RPA, and the cc-pVTZ-JKFIT
auxiliary basis for RI-J. For the SIE4x4 test set, we did not employ
RI-J, and for the G21IP test set, we did not utilise any form of RI.

Regarding the dissociation of the helium dimer, we used the aug-cc-pV6Z atomic
orbital basis along with its respective auxiliary basis for RI-J. No RI
approximation was used for the evaluation of the RPA correlation energy.
\begin{acknowledgement}
    D.~G.~acknowledges funding by the Deutsche Forschungsgemeinschaft 
    (DFG, German Research Foundation) -- 498448112. 
    D.~G.~thanks J.~Kussmann (LMU Munich) for providing a development version 
    of the FermiONs++ programme package.
\end{acknowledgement}

\begin{suppinfo}
    The complete datasets for all considered reactions, including geometries,
    reference values, and signed errors for the different methods, are readily 
    accessible at
    \url{https://github.com/dgraf-qc/c-rpa_supporting_information}.
\end{suppinfo}

\bibliography{main}

\end{document}